      [1999/12/01 v1.4c Il Nuovo Cimento]
\documentclass{cimento}

\usepackage{graphicx}
\title{GRB970228 as a prototype for short GRBs with afterglow.}
\author{M.G. Bernardini\from{1}\from{2}\ETC, 
C.L. Bianco\from{1}\from{2}, 
L. Caito\from{1}\from{2}, 
P. Chardonnet\from{1}\from{3},
A. Corsi\from{1}\from{4},
M.G. Dainotti\from{1}\from{2},
F. Fraschetti\from{1},
R. Guida\from{1}\from{2},
R. Ruffini\from{1}\from{2},
        \atque
S.S. Xue\from{1}\from{2}}
\instlist{\inst{1} ICRANet and ICRA, Piazzale della Repubblica 10, I-65100 Pescara, Italy.
           \inst{2} Dipartimento di Fisica, Universit\`a di Roma ``La Sapienza'', Piazzale Aldo Moro 5, I-00185 Roma, Italy.
           \inst{3} Universit\'e de Savoie, LAPTH-LAPP, BP 110, F-74941 Annecy-le-Vieux Cedex, France.
           \inst{4} Iasf-Roma/INAF - Via Fosso del Cavaliere 100, I-00133 Roma, Italy}
\PACSes{\PACSit{98.70.Rz}{gamma-ray sources; gamma-ray bursts}}
\begin{document}

\maketitle

\begin{abstract}
GRB970228 is analyzed as a prototype to understand the relative role of short GRBs and their associated afterglows, recently observed by Swift and HETE-II. Detailed theoretical computation of the GRB970228 light curves in selected energy bands are presented and compared with observational BeppoSAX data.
\end{abstract}

It is well known that Gamma-Ray Bursts (GRBs) are divided in two main classes on the basis of their duration with very different observational properties: ``short'' ($T_{90}< 2$ s) and ``long'' ($T_{90} > 2$ s) \cite{k93}. The existence of these two classes of bursts finds a natural explanation within our model \cite{rubr2}. In this scenario we identify short GRBs with the flash of photons emitted at the transparency, the Proper GRB (P-GRB), while the usually called long bursts coincide with the afterglow produced by the interaction of the accelerated baryons with the InterStellar Medium (ISM). The only difference among them is the value of the baryon loading $B$ which rules the amount of energy emitted in the P-GRB and the one converted into baryons kinetic energy: in the first case almost all the energy is emitted in the P-GRB, in the second one the P-GRB emission is negligible with respect to the afterglow emission and it is hardly detectable. The recent discovery of an afterglow associated to short GRBs \cite{ge05} enforces our hypothesis because it clearly represents an example for the first situation we described: the prompt emission is dominated by the P-GRB while the ``normal'' GRB emission is the following weak afterglow.

\begin{figure}
\includegraphics[width=0.5\hsize]{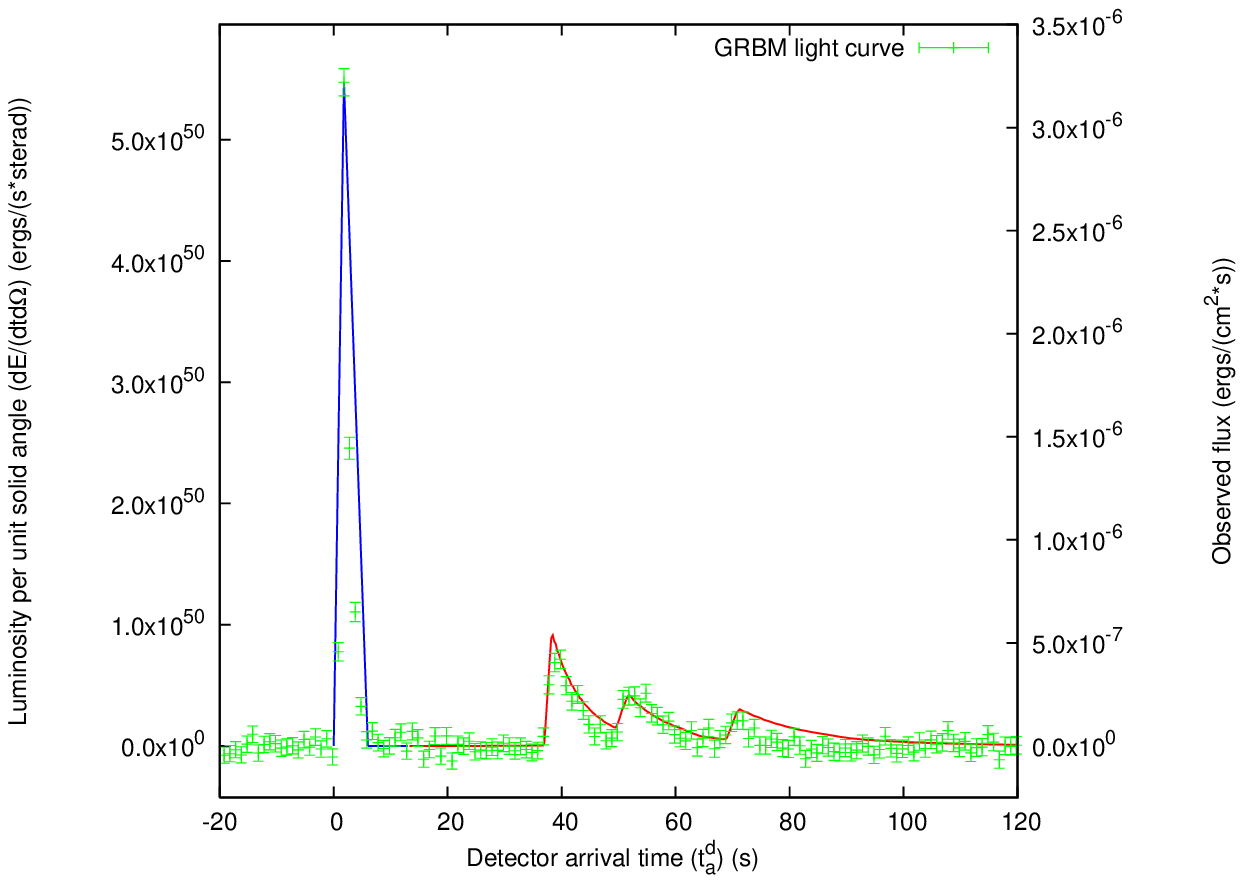}
\includegraphics[width=0.5\hsize]{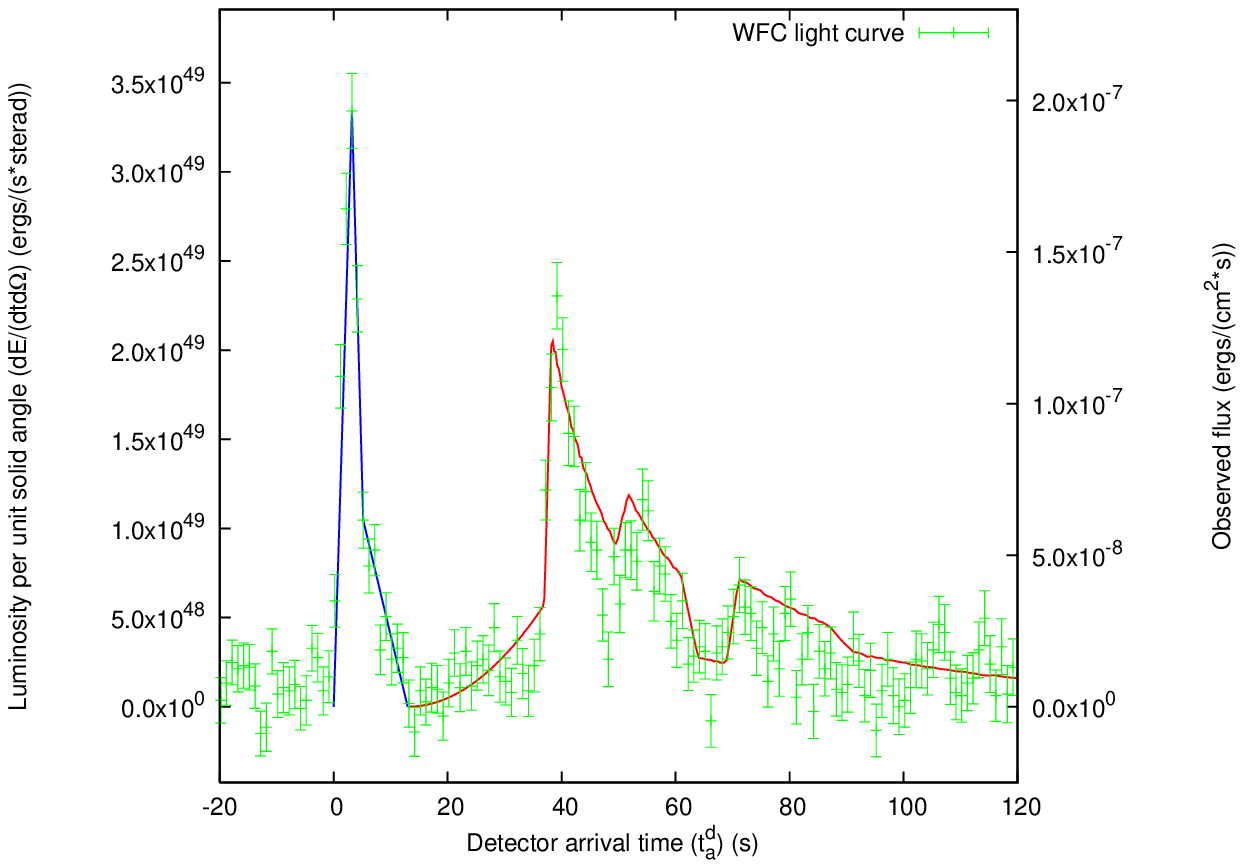}
\caption{\textit{Beppo}SAX GRBM ($40-700$ keV) and WFC ($2-26$ keV) light curves compared with our theoretical simulations of the afterglow (solid red lines). The theoretically predicted P-GRB emitted energy (whose flux is schematically represented by the solid blue line) results to be consistent with the observed one.}
\label{fig}
\end{figure}

Here we present a third situation in which, although the baryon loading $B$ is high, due to a particularly low average particle number density the afterglow peak luminosity results to be weak compared with the P-GRB. We revised our work on GRB970228 which, although it appears as a ``normal'' long GRB, could be analogous to the short ones. GRB970228 was detected by the Gamma-Ray Burst Monitor (GRBM, $40–700$ keV) and Wide Field Cameras (WFC, $2–26$ keV) on board \textit{Beppo}SAX on February $28.123620$ UT \cite{fr98}. The burst prompt emission is characterized by an initial $5$ s strong pulse followed, after $30$ s, by a set of three additional pulses of decreasing intensity \cite{fr98}. In our picture the first main pulse can be identified with the P-GRB, while the three additional pulses are an unusually low luminosity afterglow peak. This idea is enforced by the presence of a discontinuity in the spectral index between the end of the first pulse and the beginning of the others, whose spectrum appears to be more similar to the X-ray afterglow \cite{costa,fr98}. 

In Fig. \ref{fig} our theoretical fit of \textit{Beppo}SAX GRBM ($40-700$ keV) and WFC ($2-26$ keV) light curves \cite{fr98} is represented. If we interpret the first peak as the P-GRB and the three additional pulses as the afterglow peak emission, we obtain for the best fit parameters $E_{e^\pm}^{tot}=1.4\times 10^{54}$ erg and $B = 5.0\times 10^{-3}$. In order to reproduce the temporal variability of the last three pulses we determine the ISM parameters to be: $\langle n_{ism} \rangle = 9.5\times 10^{-4}$ particle/cm$^3$ and $\langle \mathcal{R} \rangle = 1.5\times 10^{-7}$. With this choice of the parameters we obtain a perfect agreement of our theoretical light curves with the observed ones and the total energy emitted in the P-GRB, $E_{P-GRB}^{tot}=1.8 \times 10^{52}$ erg, is comparable with the isotropic energy emitted in the first pulse. It is important to remark that GRB970228 is a first clear example of an intermediate situation characterized on one hand by an high baryon loading which produces a small total energy emitted in the P-GRB ($E_{P-GRB}/E_{aft}\approx 10^{-2}$), on the other hand by an extremely low ISM density that produces an afterglow with a long duration ($\approx 60$ s) and a peak luminosity which is lower than the P-GRB one.


\begin{thebibliography}{99}

\bibitem{ca05} \BY{Campana S. et al.}
  \IN{ApJ}{493}{1998}{L67}

\bibitem{costa} \BY{Costa E. et al.}
  \IN{Nature}{387}{1997}{783}
  
\bibitem{fr98} \BY{Frontera F. et al.}
  \IN{ApJ}{493}{1998}{L67}
  
\bibitem{ge05} \BY{Gehrels N. et al.}
  \IN{Nature}{437}{2005}{851}
  
\bibitem{k93} \BY{Kouveliotou C. et al.}
  \IN{ApJ}{413}{1993}{L101}
  
\bibitem{rubr2} \BY{Ruffini R., Bernardini M.G., Bianco C.L., Chardonnet P., Fraschetti F., Gurzadyan V., Vitagliano L. \atque Xue S.S.}
  \TITLE{The Blackholic energy: long and short Gamma-Ray Bursts},
  in \TITLE{COSMOLOGY AND GRAVITATION: XI$^{th}$ Brazilian School of Cosmology and Gravitation},
                  edited by \NAME{Novello M. \atque Perez Bergliaffa S.E.}
                  \IN{AIP}{782}{2005}{pp.~42-127}
                  


\end{thebibliography}
\end{document}